\documentclass[12pt]{iopart}
\usepackage{iopams}

\usepackage{graphicx}
\usepackage{dcolumn}
\usepackage{latexsym}
\usepackage{mathrsfs}
\usepackage[usenames,dvipsnames]{color}
\usepackage{hyperref}
\usepackage{color}

\usepackage[normalem]{ulem}
\def \be{ \begin{equation} } 
\def \ee{ \end{equation}} 
\def \bea{ \begin{eqnarray} } 
\def \eea{ \end{eqnarray}} 
\newcommand{\eps}{\varepsilon} 
\begin{document}
\title{Hyperuniformity in Ashkin-Teller model}
\author{Indranil Mukherjee and P. K. Mohanty}
\address{Department of Physical Sciences, Indian Institute of Science Education and Research Kolkata, Mohanpur, 741246 India.}
\ead{pkmohanty@iiserkol.ac.in}

\begin{abstract}
We  show  that equilibrium systems in $d$ dimension  that  obey the  inequality $d\nu> 2,$  known as Harris criterion, exhibit suppressed energy fluctuation  in their critical state.   Ashkin-Teller model  is  an example  in $d=2$ where   the correlation length   exponent $\nu$ varies  continuously  with  the  inter-spin interaction strength  $\lambda$   and exceeds the value $\frac d2$  set by Harris criterion  when  $\lambda$ is negative; there, the variance of 
the subsystem energy   across a length scale $l$   varies   as $l^{d-\alpha}$  with  hyperuniformity exponent $\alpha = 2(1-\nu^{-1}).$    Point  configurations constructed  by assigning  unity to  the  sites   which  has coarse-grained energy beyond a threshold value
also  exhibit  suppressed  number fluctuation  and  hyperuniformiyty with  same  exponent  $\alpha.$        
\end{abstract}\maketitle

\section{Introduction}
Hyperuniform systems are   exotic  states of matter  that lie between crystal and liquid \cite{Rev1,Rev2}.  In small length scales  they  appear disordered whereas at   large-scales  they are like perfect crystals   with   minimal or no  density fluctuations. Instances of  hyperuniformity are  observed in numerous  natural systems --arrangement of  photo receptors in birds \cite{bird1,bird2,bird3}, fish \cite{fish} and other vertebrates \cite{vert}, vegetation of ecosystems \cite{eco1,eco2}, and human settlements  \cite{Human}.  Materials like  amorphous silica \cite{silica}, superconducting vortex lattices \cite{vortex}, amorphous ice \cite{ice1} binary-mixture of  charged colloids \cite{colloid1,colloid2}  exhibit hyperuniform structure. Many  models of two-phase co-existence \cite{2phase1,2phase2,2phase3, 2phase4}, self organized  critical sates of sand-pile  models \cite{soc1,eco2}   and   other  critical  absorbing states \cite{Basu,Hexner,CLG_RO,Peter,Lebowitz} also exhibit  hyperuniformity in their critical  state.  Strangely,  patterns in prime numbers \cite{prime1,prime2}   maximally packed  extended objects \cite{pack1,pack2},   quasi-crystals  in general  \cite{quasi1,quasi2} and  several synthetic materials  \cite{matter1,matter2,matter3}  adds to  the class of  hyperuniform systems.

A general mechanism  that  leads to   stable hyperuniform structures  is yet to be found, although    some attempts have been  made in  specific  systems \cite{theory1,theory2}.    

\subsection{What is hyperuniformity}
In point configurations, hyperuniformity is defined as follows.  Let us consider  a  $L^d$  box containing $M$  particles in total,  distributed  homogeneously,  so that   the density of the system is  $\rho = M/L^d.$  Then, any subsystem of volume $V= R^d$  is expected to  have $\rho V$ number of  particles on average, but the actual number of particles  $N$ in the subsystem in any instance  would deviate from the  expected  value and it is quantified by the  variance   $\sigma^2(V) =  \langle  (N -  \rho V)^2\rangle = \langle N^2\rangle
- \langle  N\rangle^2.$   Since $N$  is a  sum  of  random variables   $s(\Delta V)$ which takes integer values $1,0$   depending on  whether a particle is present or absent  in  volume element   $\Delta V.$   The law  of large numbers then dictates  the  variance    $\sigma^2(V)$   to be   proportional to $V$ \cite{Fischer}.  This is  generically true in  many physical systems  where 
the variance of an observable  need to be additive  so that    
corresponding   susceptibility   $\sigma^2(V)/V$ in   the thermodynamic  limit   becomes a   intensive variable.  However there  are exceptions, where  $\sigma^2(V) \sim V^q$  with $q<1,$ i.e., \be \sigma^2(R)  \sim R^{d-\alpha}, \label{eq:hyp}\ee
 where $d(1-q) = \alpha.$  These systems are called  hyperuniform  as  the density fluctuations  of the  system vanishes in the $V\to\infty$ limit.  In small length scales  hyperuniform  systems  behaves like a gas (disordered), whereas   their  large-scale properties are crystal like (ordered). The hyperuniformity exponent $0<\alpha<1$   characterizes the  degree of hyperuniformity. In fact, Eq. (\ref{eq:hyp})   is formally known as Class-III  hyperuniformity (the classification of  hyperuniformity  is discussed in  Ref.  \cite{Rev1}). 

\subsection{Hyperuniform  critical states} Hyperuniformity can occur in  equilibrium systems    that  undergo a continuous phase transition  when an external  tuning parameter  crosses a critical threshold. During the phase transition, 
the  order parameter   $\phi$ of a thermodynamically large   system  vanishes  in the disordered phase and  it picks up a nonzero values in the ordered phase. At the   critical point, the system exhibits scale-invariance   as  the correlation   length   $\xi$ diverges there:  $\xi \sim |\Delta|^{-\nu},$  where $\Delta$  measures  the  deviation  of the tuning parameter  from  its  critical value ( $\Delta>0$ in ordered phase).   The mean and the variance of the order parameter scales as 
 \be 
\langle \phi \rangle \sim \Delta^\beta; ~~  \langle \phi^2 \rangle  -   \langle \phi \rangle^2 \sim |\Delta|^{-\gamma}. 
 \ee
Again, in a homogeneous and  isotropic system, the two point correlation of $\phi$  separated by a   distance  $|{\bf r}|$ is defined by 
\be
 c({\bf r} ) =  \langle \phi({\bf x}) \phi({\bf x} +{\bf r}) \rangle   - \langle \phi({\bf x}) \rangle \langle   \phi({\bf x} +{\bf r}) \rangle   \sim \frac{e^{- |{\bf r}|/\xi}}{|{\bf r}|^{d-2+\eta}}.
 \ee
 Clearly   $c({\bf r} )$ is  scale-invariant  at the  critical point   where $\xi  \to \infty.$  The set of critical  exponents $\{\nu, \beta, \gamma, \eta\}$  defines a universality class.  They  depend only  on  the dimension of the system  and the symmetry of the order parameter which is   broken at the critical point; other microscopic details  are irrelevant.  The exponents are not all independent they are  related by  scaling relations \cite{Cardy}, 
 \be
 2\beta + \gamma = d \nu; ~~  2- \eta= \frac\gamma\nu.
 \ee
Following   the  work by Ornstein-Zernike \cite{OZ1,OZ2} one can define a  total correlation function  $h({\bf r}),$ similar to that of liquids in equilibrium \cite{liq, hypercrit}, 
\be
h({\bf r} ) =c({\bf r} )+ \langle \phi \rangle \int_{L^d}  c({\bf r}-{\bf r}')   h({\bf r}') d {\bf r}. 
\ee
The first term  in the right hand side corresponds to the direct correlation function   of  observables    separated  by  distance ${\bf r}$ and   the  second  one  captures the  indirect  contribution  coming  to  these   points  through another point   at   a distance  ${\bf r}'$.
Let  the Fourier transform of $c({\bf r})$ and  $h({\bf r} )$ be  $\tilde c({\bf k})$ and  $\tilde h({\bf k} ).$ Since,  a   Fourier transform  of  the  integral   in the right hand  side  is   a  convolution,  we   get $
\tilde h({\bf k} ) = { \tilde c({\bf k})}/ {1-\langle \phi \rangle \tilde c({\bf k})}.$ The usual structure factor \cite{hypercrit} is then 
\be
S( {\bf k}) = 1+  \langle \phi \rangle  \tilde h({\bf k} ) = \frac{1}{1-\langle \phi \rangle \tilde c({\bf k})}.
\ee
At the critical point,   $c({\bf r} ) \sim  {|{\bf r}|^{-d+2-\eta}}$ and thus $c({\bf k} ) \sim - {|{\bf k}|^{\eta-2}}.$   From the scaling relations, we get  
$\eta = 2-d +  2\frac{\beta}\gamma = 2- \frac\gamma\nu;$  since critical exponents $\beta,\gamma, \nu$ are positive,   $\eta$    is  bounded    in  the range $(2-d, 2).$   Thus,  $c({\bf k})$    diverges in the  the  small $|{\bf k}|$ limit, we have 
\be 
S( {\bf k})  \sim  |{\bf k}|^\alpha ~~ {\rm with}~  \alpha = \frac\gamma\nu= 2-\eta = d- 2\frac \beta\nu.
\ee
In fact,  the above relation is  also considered as  definition of hyperuniformity 
equivalent to Eq. (\ref{eq:hyp} ) with  exponent $\alpha$ \cite{Rev1}.  A system is hyperuniform  when  the structure factor  $ S( {\bf k}) \sim |{\bf k}|^\alpha$ with  $\alpha>0.$  In effect $S( {\bf k})$ of hyperuniform systems   vanishes  as $|{\bf k}|\to 0.$  Based on this,  hyperuniformity is classified  into three  categories -
\begin{eqnarray}
{\bf Class ~ I}~ (\alpha>1) &:&  \sigma^2(R)  \sim R^{d-1}\cr
{\bf Class ~ II}~  (\alpha=1)&:& \sigma^2(R)  \sim R^{d-1} \ln(R)\cr
{\bf Class  ~III}~  (0<\alpha<1)&:&  \sigma^2(R)  \sim R^{d-\alpha}
\end{eqnarray}
Hyperuniform critical states  generally   belong  to {\bf Class  ~III} \cite{Rev1},  the weakest among the three classes.

 Hyperuniformity is  not  restricted to  equilibrium.  Recent  works indicate that, under nonequilibrium conditions, two-dimensional crystals exhibit hyperuniformity \cite{2DCryst}.  In  absorbing phase transition,  the sub-critical  phase is non-fluctuating.  The  system  approach  the absorbing  state   by  pushing particles  from active local regions to  their neighbourhood, thereby  reducing  density fluctuations - or  creating hyperuniformity locally.   This   process happens over a  finite distance (the correlation length $\xi$)  and  it extends  all over the system as  the critical  point is approached ($\xi\to \infty$). In fact, critical  absorbing states are generically hyperuniform \cite{Hexner}.  In Manna sand-pile model,  at the critical density $\rho_c$ the density fluctuation  $\sigma^2(l)  = \langle \rho( l)^2 \rangle  -  \langle\rho( l)\rangle^2$ in a volume $V= l^d$  is found to be, 
\be
\sigma^2(l)  = \langle \rho( l)^2 \rangle  -  \langle\rho( l)\rangle^2 \sim l ^{d-\alpha} \label{eq:hyp_rho}
\ee
with $\alpha=  0.425, 0.45, 0.24$ respectively for $d=1,2,3$ \cite{Hexner}.   Direct measurement of  structure factor also  confirm  that $S(k) \sim k^\alpha$ with  same  $\alpha$ value. This work by Hexner and Levin   \cite{Hexner} also    showed that, within the acceptable error limits,  the hyperuniformity exponent  $\alpha$  matches with  $2-\eta,$ where $\eta$ is  the  exponent  related to the  correlation function  of activity density $\rho_a$  at  the critical point, i.e., $
 \langle \rho_a({\bf x}) \rho_a({\bf x} +{\bf r}) \rangle   - \langle \rho_a({\bf x}) \rangle \langle   \rho_a({\bf x} +{\bf r}) \rangle   \sim {|{\bf r}|^{-d+2-\eta}}.$ In fact, an  exceptionally   uniform  density profile was observed earlier  \cite{Basu}  in one dimensional  Manna models \cite{Manna}, though it was not named as hyperuniformity; the  exponent $\alpha \equiv  d- 2 \frac\beta\nu =0.496$ (calculated using  $\beta, \gamma$ as that of directed percolation \cite{DP})   is quite close to $\alpha=  0.425$  obtained  in Ref. \cite{Hexner}. The discrepancy  owes  to   the fact that  (a) there are   strong   finite size effects  and (b) the  system takes  unusually long time to reach its correct steady state  where   density  profile   is  hyperuniform \cite{Basu}.

 Other  models  having   absorbing phase transition  like    Oslo model \cite{Oslo1,Oslo2},  conserved lattice gas  \cite{CLG},   random organization (RO)  models \cite{RO}, and facilitated  exclusion process \cite{Urna} also exhibit hyperuniformity, which is reported in  Refs. \cite{Peter} , \cite{CLG_RO}, and \cite{Lebowitz} respectively. For a recent review   on  hyperuniformity in  non-equilibrium systems, see Ref. \cite{Rev_Hyp_Noneq}.
 
\subsection{Objective of our study}
 In this  article we explore another  possibility where internal energy  of the system, at its critical point,  exhibit hyperuniform distribution.
 In 2$^{nd}$ order  equilibrium phase transitions,   the average of energy  of the system vary continuously across  the critical point but the  variance,  known as the specific heat $C_v$ diverges, 
\begin{equation}\label{eq:alpha}
C_v= \frac{1}{V} (\langle E^{2} \rangle - \langle E)\rangle^{2} \sim \Delta ^{-a}.
\end{equation}
Here  we denote the specific heat exponent as $a$ (as hyperuniformity exponent  is usually    denoted as  $\alpha$ in recent literature). 
In a finite system, the correlation length $\xi$ is  limited  by the system size  $L$ and  thus $\Delta \sim L^{-1/\nu},$  resulting in 
\be
\sigma_E^2(L) = \langle E^{2} \rangle - \langle E\rangle^{2} \sim L^{d +\frac{a}\nu}.
\ee
Energy $E_l$   of a  large subsystem  of size $l\times l,$  with $1 \ll l \ll L$ is  expected to show a  similar scaling behaviour, 
\be
\sigma_{E_l}^2(l) = \langle E_l^{2} \rangle - \langle E_l \rangle^{2} \sim l^{d +\frac{a}\nu}.
\ee
A comparison with Eq. (\ref{eq:hyp})  gives us  $\alpha = -\frac a \nu.$  Since hyperuniformity   distribution is possible  only for $0<\alpha <1,$  we need $-1 <\frac a \nu<0.$  Is  it possible ?  Usually,  the specific heat  of a system diverges 
 at the  critical point, which corresponds to  $a>0$ and  $a$ is positive. We  consider  Ashkin-Teller model in 2d as an example, where  $a$  can become  negative  in some regime  of the parameter space. 
 There we show that, energy  of a subsystem  exhibit hyperuniformity  with   
 \be 
 \alpha = \frac{|a|}{\nu}=2(1- \nu^{-1}), \label{eq:hyp_nu}
 \ee
where in the last step we use the scaling relation $2-a = d \nu.$  
 \begin{figure}[t]
\vspace*{.5 cm}
\centering
\includegraphics[width=7cm]{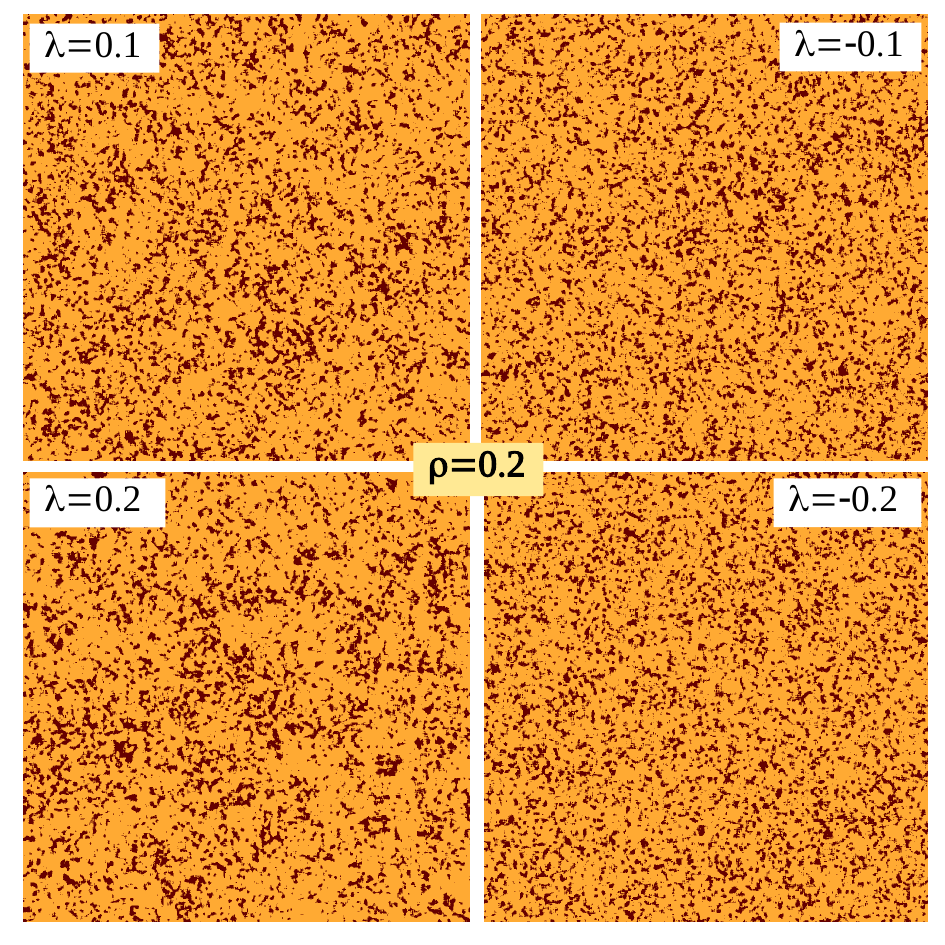} \includegraphics[width=7cm]{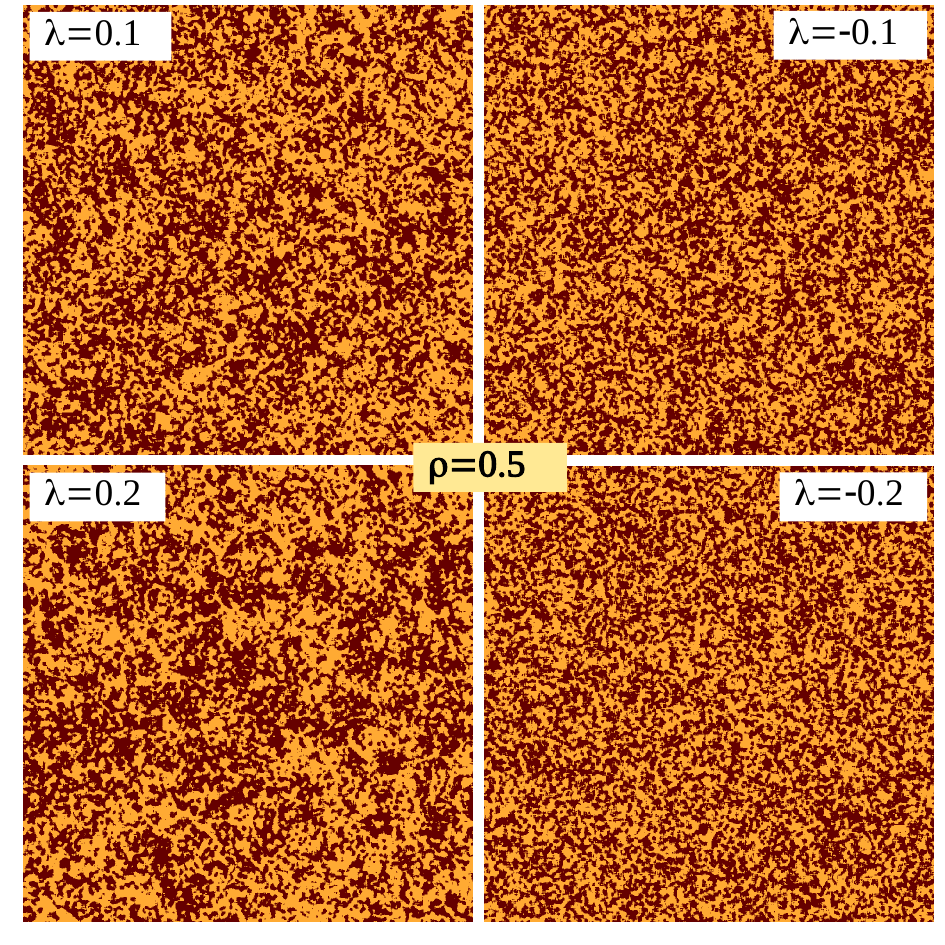} 
\caption{Point configurations  generated   in Ashkin Teller model by assigning, $1$(dark colour) to sites  whose energy exceeds  a  pre-determined  threshold, else $0$ (light colour) on a  $1024\times 1024$  square lattice. Configurations  with negative $\lambda$  (hyperuniform) are compared  with those  having  positive values (uniform) for  particle density  is $\rho=0.2$ (left), 0.5 (right). }
\label{fig:lambda_plots}
\end{figure}

Hyperuniformity is primarily studied in systems  which are described by point configurations.  To substantiate,   we  use a  threshold  filter  that   converts the real variable $E_l$  to integers $1,0$ representing presence  and absence of particles, and  thereby generates  point configurations. Although
mean density  of these point configurations depend on the  threshold, we find  that   the fluctuations  exhibit  hyperuniformity  with  exponent $\alpha$  that does not depend on  particle density. $\alpha$ is given by  Eq. (\ref{eq:hyp_nu}). In Fig. \ref{fig:lambda_plots} we    represent   the point configurations   of  the  Ashkin-Teller model pictorially for    interaction strengths $\lambda=\pm 0.1, \pm 0.2$   and two different  particle densities $\rho=0.2$(left), $0.5$ (right).
The  dark colour   there  represent occupied sites.  Clearly, configurations with negative $\lambda$  appears  smoother.  Details  are  discussed   in  the following section.

 \section{Ashkin-Teller  Model}
Ashkin Teller (AT) model \cite{AT_1943, Wu_Lin, Kadanoff_1977, Baxter} is a two-layer Ising system  defined  on a  $L\times L$ square lattice with periodic boundary conditions in both directions.  The sites of the lattice   with coordinates $(x,y)$  
is represented  by vectors ${\bf i} \equiv (x,y)$ where $x,y=1,2,\dots, L.$    
Each site ${\bf i}$ of the lattice carries two different Ising spins $\sigma_{\bf i}=\pm$ and $\tau_{\bf i}=\pm$ which  interact as follows.
\begin{equation} \label{eq:AT_H}
 H = -J\sum_{\langle {\bf ij}\rangle} \sigma_{\bf i} \sigma_{\bf j} - J \sum_{\langle {\bf ij}\rangle}\tau_{\bf i} \tau_{\bf j} - \lambda \sum_{\langle {\bf ij}\rangle}\sigma_{\bf i} \sigma_{\bf j} \tau_{\bf i} \tau_{\bf j}. 
\end{equation}
Here $\langle {\bf ij}\rangle$ denotes a pair of nearest-neighbor sites, $J$ is  the strength of  intra-spin interactions and $\lambda$ represents interactions among  $\sigma$ and $\tau$ spins. The  energy of the   system in   Eq. (\ref{eq:AT_H}) remains unchanged   under one or  more of the   the following transformations:
\be
\sigma \rightarrow -\sigma;  ~~\tau \rightarrow -\tau;~~ \sigma \rightarrow \tau. 
\ee
Thus  it is natural to   consider  that the  quantities  $M=\langle\sigma\rangle= \langle\tau\rangle,$  and  $P=\langle\sigma\tau\rangle$  which obey these symmetries may  break giving rise to an ordered phase.  Formally these observable are  referred to as  magnetisation and polarisation respectively.  In AT model, both $M$ and  $P$  break the  symmetry across the self-dual critical line \cite{Wegner},
\be \label{eq:TcJclam}
T_{c} =1, \lambda_c=\lambda, J_c = \frac12 \sinh^{-1}(e^{-2\lambda})
\ee
and  generate   finite magnetization  ($M\ne0$)  and  finite polarization ($P\ne0$) simultaneously.
The  electric and magnetic phase transitions are characterized by  the respective  critical exponents (denoted by subscripts $e,m$ respectively)
\begin{eqnarray}
 \label{eq:beta_mag}
 \langle M \rangle \sim \Delta ^{\beta_m} ; ~~ \chi_m = \langle M^{2} \rangle - \langle M \rangle^{2} \sim \Delta ^{-\gamma_m}; ~~ \cr
 ~~ \langle P \rangle\sim \Delta ^{\beta_e};~~ \chi_e = \langle P^{2} \rangle - \langle P \rangle^{2} \sim\Delta ^{-\gamma_e};
\end{eqnarray}
where $\Delta=T_{c}-T.$   
 The  other critical exponents  $\nu,a$  associated with the  correlation length  and  the specific heat  respectively,   do  not  depend on the  order parameters  and  thus carry  no  subscripts,
\begin{equation}\label{eq:alpha}
 \xi  \sim \Delta ^{-\nu}; ~~~C_v= \frac{1}{L^2} (\langle E^{2} \rangle - \langle E)\rangle^{2} \sim \Delta ^{-a}.
\end{equation}

The  Ashkin-Teller model can be mapped to  the eight vortex model introduced and solved by Baxter \cite{ Baxter3, Baxter}.  This  mapping  and re-normalization group arguments \cite{Wu_Lin} provide exact value of the critical exponents in terms of $\mu = cos^{-1} \left(  e^{2\lambda} \sinh(2\lambda)    \right)$,
\be
\nu=\frac{2 (\mu-\pi)}{4\mu-3\pi}; ~ a=2(1-\nu); \beta_e= \frac{2 \nu -1}{4}; ~\gamma_e =\frac{1}{2}+\nu; \beta_m=\frac{\nu}{8};~\gamma_m =\frac{7\nu}{4}. \label{eq:exact_mag_AT}
\ee
Note  that  the critical exponents  of magnetic transition satisfy the following relations: $\frac{\beta_m}{\nu} = \beta_m^0,\frac{\gamma_m}{\nu} =  \gamma_m^0,$  where $\{ \beta_m^0=\frac18, \gamma_m^0= \frac74\}$ are the critical exponents of Ising universality class in $d=2$. This is the well-known weak universality  scenario proposed by Suzuki \cite{Suzuki}, and observed in several experiments \cite{Guggenheim, Back}. According to  weak-universality hypothesis,  universality  feature  of a  phase transition   should be determined  on how the physical quantities  depend   on  the  correlation length  $\xi$ which is an emerging property of of the system,   not on   $\Delta=T_c-T$    because   $T_c$  is  a non-universal  -- it depends on the  microscopic details of the system.

The electric transition  violates both universality and weak-universality hypothesis; all its critical exponents vary continuously with $\lambda.$ Recent works  propose a  super universality hypothesis  \cite{suh} which states the following. Critical behaviour of a system 
form a  super-universality class when  the  scaling function that relates the  binder cumulant  of the system  with the  second moment correlation length in units of  system size $L$ remains invariant along the  critical line, even when  the  critical exponents  vary.
 It turns out that both magnetic and electric  phase transition of  AT model   belong to Ising-super-universality class \cite{suh}.

\subsection{Energy distribution }

\begin{figure}[t]
\centering
\includegraphics[width=6.25cm]{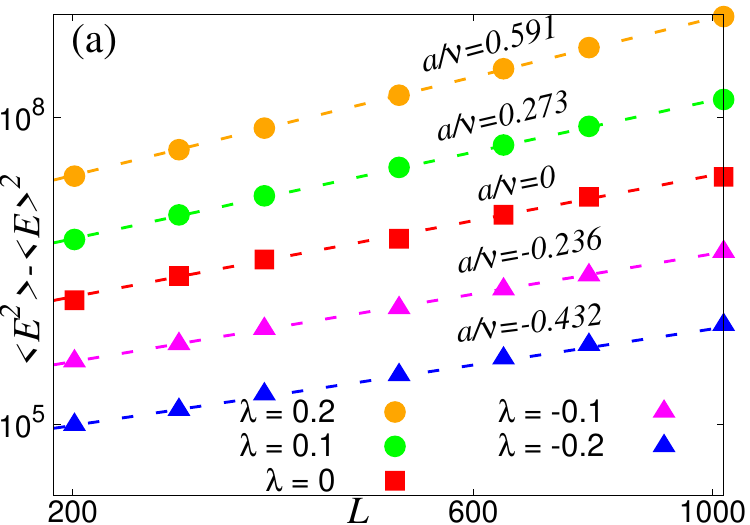}\hspace{.7cm}
\includegraphics[width=6.25cm]{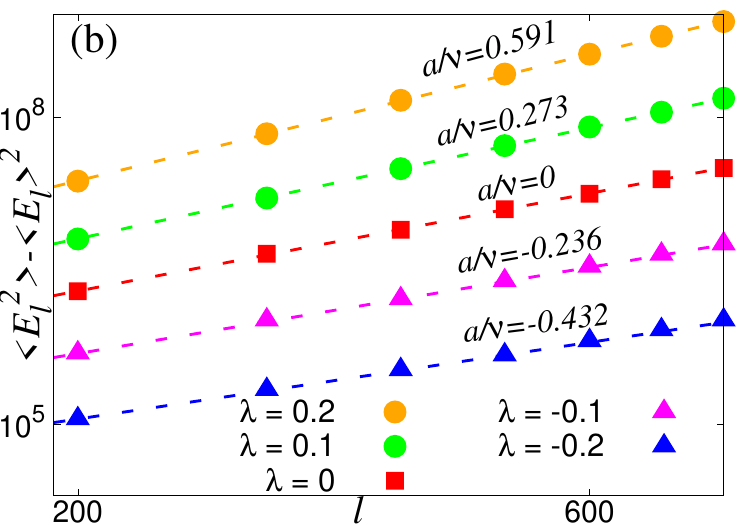}
\caption{(a) Log-scale plot of  variance  of total energy $\langle E^{2} \rangle - \langle E\rangle^{2}$ as a function of system size $L$ for different $\lambda$ values (symbols) are compared with  $L^{d+\frac{a}{\nu}}$ (dashed lines).  (b) Energy fluctuations in $l\times l$ subsystems: $\langle E^{2}_l \rangle - \langle E_l\rangle^{2}$ as a function of  $l$ in log-scale for different $\lambda$ values (symbols)  matches with  dashed lines  $l^{d+\frac{a}{\nu}}$   where $a$ is given by  Eq. (\ref{eq:exact_mag_AT}). 
In all cases $L=1024,$  and 
averaging is done  over more than $10^6$ samples; data are also scaled by  factors  $\{\frac12,1,2,4,8\}$ (bottom to top) for better visibility. }
\label{fig:Cv}
\end{figure}

 In this  article we  focus on  studying  energy distribution   along the  critical line.  In a finite system,  the  correlation length $\xi$, which  is supposed  to diverge as $\Delta^{-\nu},$   is   bounded by the   system size $L$ and  thus  $L\equiv \Delta^{-\nu}.$   Then, 
 \be 
 \langle E^{2} \rangle - \langle E\rangle^{2} \sim L^d \Delta ^{-\alpha}=L^{d+\frac a\nu}.
 \label{eq:E2}
 \ee
In Fig.  \ref{fig:Cv}(a) we have plotted $\langle E^{2} \rangle - \langle E\rangle^{2}$ as a function of system size $L$  in log-scale, for different $\lambda$; for comparison,  dashed lines with slope  $d+ \frac{a}{\nu}$   are drawn  along the  symbols.  It  is no surprise that they agree well.

We now look at  the  energy of the subsystems of size $l\times l,$   where  $l\ll L.$ First  let us define  the  local energy at each site, 
\bea 
H_{\bf i} &\equiv& H_{x,y} = -J s_{x,y} \left( s_{x,y+1} + s_{x+1,y}\right)  -J \tau_{x,y} \left( \tau_{x,y+1} + \tau_{x+1,y}\right)\cr  &&-\lambda \left(  s_{x,y}\tau_{x,y}  (s_{x,y+1} \tau_{x,y+1} +s_{x+1,y}\tau_{x+1,y} \right)
\label{eq:Hxy}
\eea
so that the total energy  given  in Eq. (\ref{eq:AT_H})  $H = \sum_{\bf i} H_{\bf i}.$ The homogeneity of  the system  states  that  $\langle H \rangle = L^2 \langle H_{\bf i} \rangle.$ Energy of a subsystem of size $l\times l$ is defined as $E_l = \sum\limits_{ {\bf i} \in l \times l} H_{\bf i},$  where the $l\times l$ square  is placed   on the   lattice   such that 
its sides are aligned and   one of its corner  matched   with  a randomly chosen site ${\bf j}.$  Clearly, $\langle E_l\rangle = l^2 \langle H_{\bf i} \rangle.$   The variance  
of $E_l,$  $\sigma^2_{E_l}(l)= \langle E_l^2 \rangle -  \langle E_l \rangle^2,$  is    calculated  from Monte Carlo simulations  when the system is in steady state.  In Fig.  \ref{fig:Cv}(b)  we plot  $\sigma^2_{E_l}(l)$  as a  function of $l$ in log-scale  for different $\lambda.$  Along with  the data (symbols), for comparison,  we draw dashed lines of   slope $d+\frac a\nu.$  A good match suggests that the  subsystem  energy fluctuation also obey the  scaling $\sigma^2_{E_l}(l) \sim l^{d+\frac a \nu},$  similar to Eq.  (\ref{eq:E2}). Thus, the  variance of  subsystem  energy  also exhibit   hyperuniformity   when $a <0,$  i.e. when  $\lambda <0.$

\begin{figure}[t]
\centering
\includegraphics[width=5.12cm]{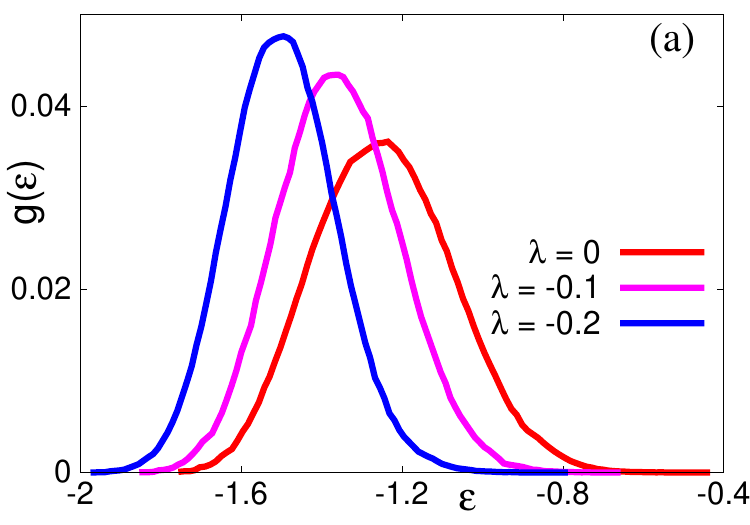}
\includegraphics[width=5.12cm]{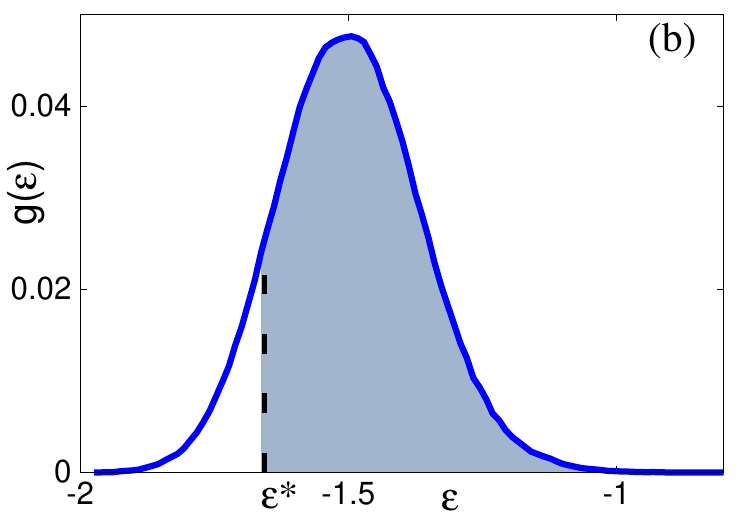}
\includegraphics[width=5.12cm]{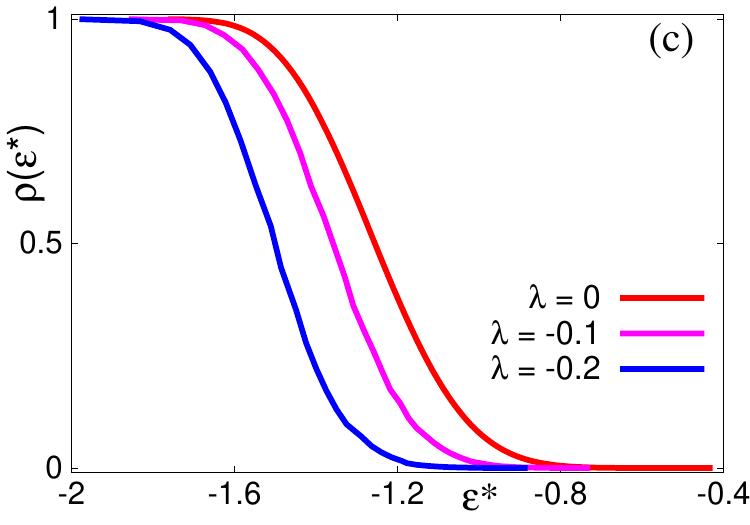}
\caption{(a) The distribution $g(\varepsilon)$ of coarse-grained local energy $\varepsilon$ defined in Eq. (\ref{eq:vareps}) with  $l=8.$ Different curves corresponds to  $\lambda = -0.2, -0.1, 0$. (b) Schematic representation  of particle density $\rho(\eps^*),$ as in Eq. (\ref{eq:rhoeps}). The shaded  represents   particle density. 
(c) $\rho(\varepsilon^*)$ as a function of the  threshold $\varepsilon^*$ for $\lambda = -0.2, -0.1, 0$.}
\label{fig:P(e)}
\end{figure}

\subsection{Hyperuniformity in point configurations}
 Hyperuniformity is generally defined on  point configurations  carrying discrete variables  on  lattice or in continuum.  Their  pictorial representations are visually appealing as discrete  variables generates sharp contrast --  hyperuniform   systems  appears strikingly different and more uniform  than   their  counterparts  where    particles are distributed   randomly.  Local energy $H_{\bf i},$ being a real variable, cannot display such a contrast. 
 In the following we try to construct  point configurations that faithfully represent  the local energy. 
 This can be done using a threshold energy $\varepsilon^*$ that places a point (particle) at every site $(x, y)$ with energy $H_{x,y}$ larger than $\varepsilon^*$; the particle density $\rho$ of the point configurations can be controlled by varying $\varepsilon^*$. This process does generate point configurations, but $\rho(\varepsilon^*)$ is limited to certain fixed values. This limitation arises because the local energy $H_{x,y}$ in Eq. (\ref{eq:Hxy}) is a discrete set (sum of integer multiples of $J$ and $\lambda$). To achieve a continuous variation of $\rho$, we need a coarse-grained energy at each site.

Let $l$ be the length scale over which coarsening is performed. We define $\varepsilon_{\bf i} \equiv \varepsilon_{x,y}$ as the average energy of all the sites belonging to an $l \times l$ square centered at site ${\bf i} \equiv (x, y)$,

\be
\varepsilon_{\bf i} = \frac{1}{l^2} \sum_{m,n=0}^{l-1} H_{x+m,y+n}.\label{eq:vareps}
\ee

The energy spectrum generated by $\varepsilon_i$ is bounded with an energy gap (difference of consecutive energy values) of $ {\cal O} \left(\frac{1}{l^2}\right)$. The energy spectrum becomes continuous when $l$ is large.
 For our purpose $l=8$ turns out  right.  First we calculate the  probability density function  $g(\varepsilon)$ for $\lambda=0,-0.1, -0.2,$   which is shown in Fig. \ref{fig:P(e)} (a).  Further we define $n_{\bf  i} = \theta(\eps_{\bf  i} -\eps^*)$  where  $\theta(x)$ is  the  Heaviside step function that  takes   the value $0$ when $x$ is negative, and $1$ otherwise.
  Thus,  $n_{\bf  i}$    behaves like   the  presence and absence of a particle  at site ${\bf i}.$   The  threshold  $\eps^*$  controls the density of the  particles 
\be 
\rho(\eps^*) =\frac{1}{L^2}\sum_{\bf i} n_i = \int_{\eps^*}^\infty g(\eps) d\eps.
\label{eq:rhoeps}
\ee
This is demonstrated  in Fig. \ref{fig:P(e)} (b) for $\lambda=-0.2;$ the shaded area there  represents   the  particle density $\rho(\eps^*).$   A plot of $\rho(\eps^*)$ is  shown in  Fig. \ref{fig:P(e)} (c) for different $\lambda.$

\begin{figure}[t]
\centering
\includegraphics[width=6.25cm]{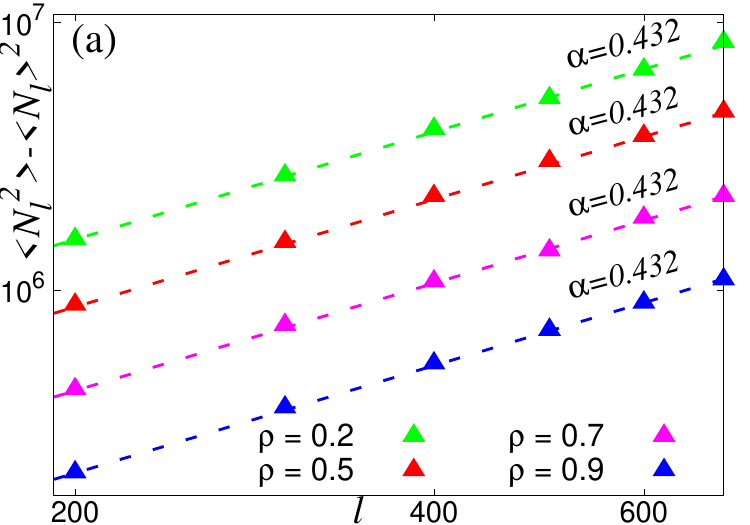}\hspace{.7cm}
\includegraphics[width=6.25cm]{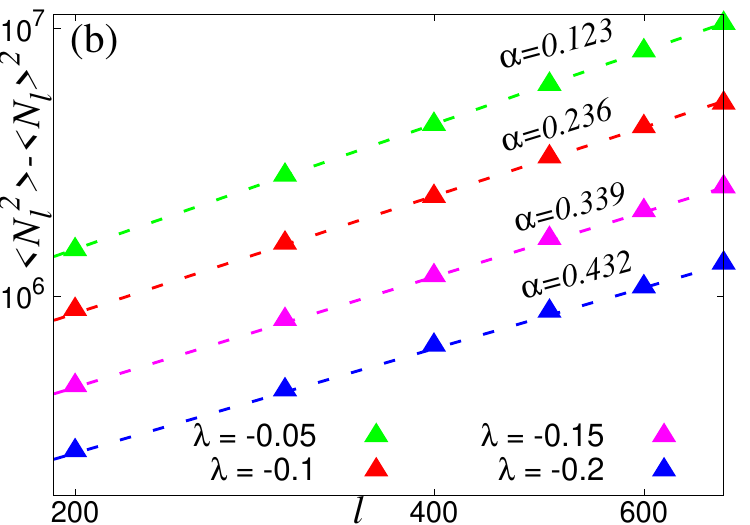}
\caption{Log-scale plot of number fluctuations in $l\times l$ subsystems  $\langle N_l^{2} \rangle - \langle N_l \rangle^{2}$ as a function of $l.$  
(a) $\lambda = -0.2$ and densities   $\rho=0.2,0.5,0.7,0.9.$ Dashed lines  corresponds to $l^{d-\alpha}$ with $\alpha= \frac{|a|}{\nu}=0.432,$ known exactly from Eq. (\ref{eq:exact_mag_AT}).
 (b)  $\rho=0.5$ and $\lambda=-0.05,-0.1,-0.15,-0.2.$ Data (in symbols) are compared with the exact values (dashed-lines) known   from  Eq. (\ref{eq:exact_mag_AT}). In all cases, $L=1024,$ and averaging is done  for $10^4$ or more samples; data  are scaled by  factors $\{ \frac14,\frac12,1,2\}$  (bottom to top) for better visibility.}
\label{fig:Nl}
\end{figure}

We are now ready to ask, if the  number fluctuations in  these  point configurations exhibits hyperuniformity. 
Number of particles in a subsystem   of size $l\times l$ is   $N_l = \sum\limits_{ {\bf i} \in l \times l} n_{\bf i}.$   By construction,  $\langle  N_l \rangle = l^2  \rho(\eps^*).$  We want to   calculate the   variance,  
\be
\sigma_{N_l}^2(l) = \langle N_l^{2} \rangle - \langle N_l \rangle^{2} \sim l^{d -\alpha}.
\ee
and find out if   it scales as  $l^{d -\alpha}$   with  $\alpha$ in the range $(0,1).$  In  Fig. \ref{fig:Nl} (a) we plot $N_l$ vs $l$   for a  fixed $\lambda=-0.2$  and  different densities $\rho.$ All of them show  the same power-law  scaling  with $\alpha= \frac{|a|}\nu =0.432$  (dashed lines), calculated  from  Eq. (\ref{eq:exact_mag_AT}).  Figure  \ref{fig:Nl}(b)   shows  $\sigma_{N_l}^2(l)$  in log-scale for different values of $\lambda$  and they  exhibit different  hyperuniformity exponents, consistent with   known values  $\alpha =\frac{|a|}\nu.$

\begin{figure}[h]
\vspace*{.5 cm}
\centering
\includegraphics[width=5.12cm]{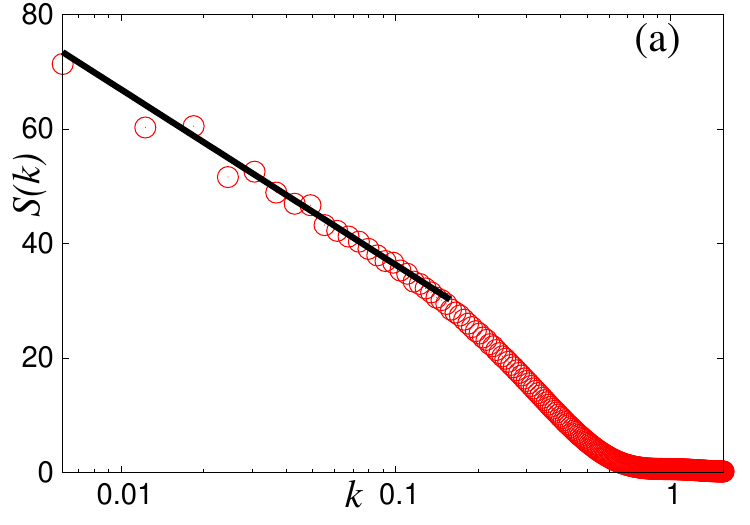}
\includegraphics[width=5.12cm]{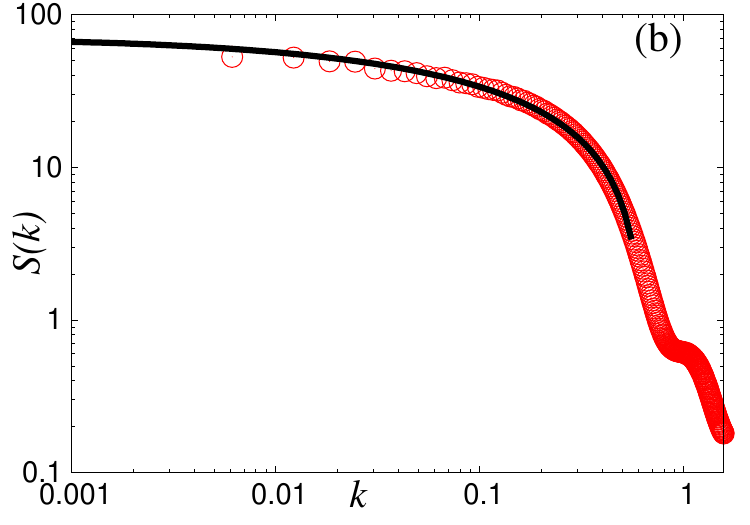}
\includegraphics[width=5.12cm]{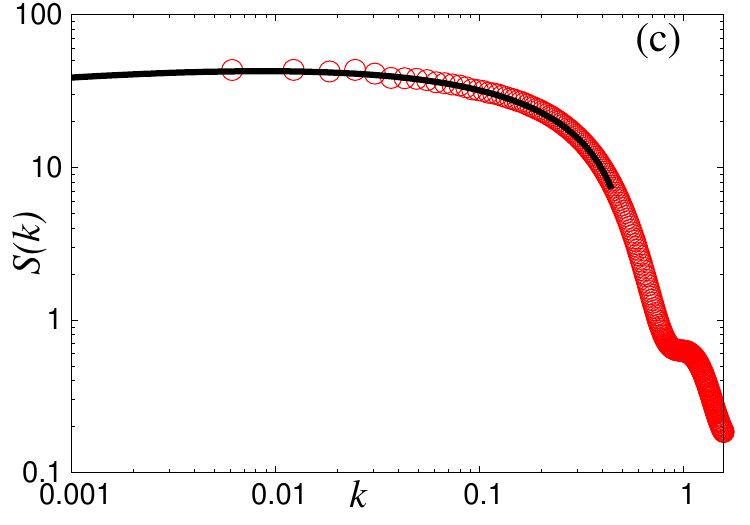}
\caption{ Radial structure factor $S(k)$ as a function of $k$ for different $\lambda$ values (symbols)  are fitted   to  the form $-A k^{\alpha} \ln(B k)$ shown as lines. (a) $\lambda=0$ ($\alpha=0$):  $S(k)$ is plotted  with  $x$-axis in log-scale  to see a linear behaviour  for small $k$  that corresponds to  $S(k) \sim \ln(k).$  (b)  $\lambda=-0.05$ ($\alpha=0.123$) and (c) $\lambda=-0.1$ ($\alpha=0.236$):  $S(k)$ is shown in log-log scale. In all cases $\eps^*$ is chosen such that $\rho=\frac12,$ and $S(k)$  is averaged  $10^5$ point configurations.}
\label{fig:s(k)}
\end{figure}

\subsection{The structure factor}

The structure factor of point configurations generated from the steady-state energy profile of the AT model can be calculated as follows. Given that the exponent $\alpha$ is independent of particle density, we can choose any particle density. For our calculations, we consider point configurations with a particle density $\rho(\varepsilon^*) = \frac{1}{2}$ (or  $N= \frac{L^2}{2}$)  by selecting a suitable $\varepsilon^*$. Let $\mathbf{R}_j$ be the position of the $j$-th particle, where $j = 1, 2, \ldots, N$. The structure factor $S(\mathbf{k})$ for the point configuration $\{\mathbf{R}_j\}$ is given by:
\be
S(\mathbf{k}) = \left\langle \left| \sum_{j=1}^{N} e^{i \mathbf{R}_j \cdot \mathbf{k}} \right|^2 \right\rangle = \left\langle \left| \sum_{i \in L \times L} n(\mathbf{r}_i) e^{i \mathbf{r}_i \cdot \mathbf{k}} \right|^2 \right\rangle,
\ee
where $\langle \cdot \rangle$ represents the statistical average over the configurations, and $\mathbf{k} \equiv (k_x, k_y) = \frac{2\pi}{L} (n_x, n_y)$ with $n_x, n_y = 1, 2, \ldots, L$. The radial form of the structure factor is then:

\be
S(k) = \frac{\sum_{\mathbf{k}} S(\mathbf{k}) \delta(k - |\mathbf{k}|)}{\sum_{\mathbf{k}} \delta(k - |\mathbf{k}|)}.
\ee

For the AT model, point configurations at $\lambda = -0.05$ and $\lambda = -0.1$ are expected to have a radial structure factor $S(k) \sim k^\alpha$ with $\alpha = 0.123$ and $\alpha = 0.236$, respectively, as $k \to 0$. However, it is well known that the structure factor of critical systems in two dimensions does not exhibit pure power laws but contains a multiplicative logarithmic factor, i.e., $S(k) \sim k^\alpha \ln k$ \cite{Rev1}. The presence of a strong logarithmic correction can be verified in point configurations of the AT model at $\lambda = 0$, where $\alpha = 0$ and thus $S(k) \sim \ln(k)$.

In Fig.  \ref{fig:s(k)}, we have plotted the radial structure factor $S(k)$ for $\lambda = 0, -0.05, -0.1$. In Fig.  \ref{fig:s(k)}(a), the plot of $S(k)$ on a semi-log scale exhibits linearity for small $k$, confirming the existence of a strong logarithmic correction. Figures  \ref{fig:s(k)} (b) and 5(c) show the same for $\lambda = -0.05$ and $\lambda = -0.1$ on a log-log scale. For small $k$, the data (symbols) do not show a straight line with a positive slope $\alpha$; the power-law is masked by a strong logarithmic correction. However, a function of the form $-A k^\alpha \ln(Bk)$ (black line) fits the data quite well with $\alpha = 0.123$ and $\alpha = 0.236$, respectively, along with suitable choices of constants $A$ and $B$.

 \section{Conclusion and Discussion}
 In this article  we demonstrate that  Ashkin-Teller model  on a two dimensional $(d=2)$ square lattice exhibit hyperuniform   energy distribution  when   the  inter-spin interaction  $\lambda$  becomes negative: variance of  energy of the  subsystems of  size $l \times l$  is proportional to $ l^{d -\alpha}$  with $\alpha = \frac{|a|}\nu= d- \frac2\nu$ where $\nu, a$  are   the correlation length and 
 specific heat exponents respectively.  Any point  configuration constructed by assigning unity to sites  having   energy  (coarse-grained) exceeding  beyond a  threshold value,  leads to suppressed  number  fluctuations  $l^{-\alpha}$  across a length scale $l,$  which  vanishes  as $l \to \infty.$ The exponent $\alpha= d- \frac2\nu$  does not depend on the  threshold value  of energy  which   controls the  particle density  of  the point configurations.

Besides Ashkin-Teller model several other  spin systems  exhibit  reduced specific heat  at  the  their magnetic transition point, as their  specific heat exponent  $a$  is negative. Some examples  include   XY \cite{3D_XY}  and Heisenberg  model \cite{3D_O3}  in  $d=3$,    random-field Ising models in $d=2$ \cite{alpha-I}, and $d=3$ \cite{alpha-II}. There too, one expects to see  hyperuniform energy distribution.  The advantage in AT model is that   here   $\alpha$ varies continuously.  This enables  us to  utilise the 
  steady-state energy configurations   of  critical  Ashkin-teller model  to  generate real-valued hyperuniform distributions   on  two dimensional lattices. Using a threshold on a coarse-grained   local energy  one can also  generate  hyperuniform   particle distribution with  any desired  exponent and density.

Equilibrium systems exhibit hyperuniform energy distribution when $0<\alpha < 1$ or equivalently when  $d\nu>2.$ This condition  is the well-known  Harris  criterion \cite{Harris} which states that  continuous  phase transitions with $d\nu>2$ are stable against quenched   spatial randomness.  This inequality is  derived later  for  systems with uncorrelated   point  disorder \cite{Chayes}.  There are numerous models   that obey  Harris criteria  applies \cite{Brooks}, but exceptions  do exist  \cite{Sing,Stinchcombe, Goth, Fayfar}.   Since  systems  obeying  Harris criteria   exhibit    hyperuniform  energy distribution, one may conjecture that quenched   spatial disorder  is  irrelevant   in  systems  where  energy distribution is  hyperuniform. Further study is required  to establish if such a conjecture is indeed true.

\section*{Aknowledgement}  
The authors would like to acknowledge helpful discussions with Aikya Banerjee and extend their gratitude to Soumya K Saha for assistance in generating figure \ref{fig:s(k)}. PKM gratefully acknowledges the financial support provided by the Science and Engineering Research Board (SERB), Department of Science and Technology, Government of India, under project reference no. MTR/2023/000644 for conducting this research.

\end{document}